\documentclass[useAMS,usenatbib]{mn2e}
\usepackage{epsfig}
\usepackage{amssymb}
\usepackage{amsmath}

\newcommand\ion[2]{#1\,{\scshape{#2}}}%                       % ion, i.e., CII = \ion{C}{ii}

\title[Broad-line-region hydrogen line ratios, AGN reddenings, and accretion disc sizes] {The case for cases B and C: intrinsic hydrogen line ratios of the broad-line region of active galactic nuclei, reddenings, and accretion disc sizes}
\author[C. M. Gaskell]{C. Martin Gaskell\thanks{E-mail:
mgaskell@ucsc.edu} \\
\\Department of Astronomy and Astrophysics, University of California,
Santa Cruz, CA 95064\\}
\begin{document}

\date{}

\pagerange{\pageref{firstpage}--\pageref{lastpage}} \pubyear{2015}

\maketitle

\label{firstpage}

\begin{abstract}
\\
Low-redshift active galactic nuclei (AGNs) with extremely blue optical spectral indices are shown to have a mean, velocity-averaged, broad-line H$\alpha$/H$\beta$ ratio of $\thickapprox 2.72 \pm 0.04$, consistent with a Baker-Menzel Case B value. Comparison of a wide range of properties of the very bluest AGNs with those of a luminosity-matched subset of the Dong et al.\@ blue AGN sample indicates that the only difference is the internal reddening. Ultraviolet fluxes are brighter for the bluest AGNs by an amount consistent with the flat AGN reddening curve of Gaskell et al.\@ (2004).  The lack of a significant difference in the {\it GALEX} (FUV--NUV) colour index strongly rules out a steep SMC-like reddening curve and also argues against an intrinsically harder spectrum for the bluest AGNs.  For very blue AGNs the Ly$\alpha$/H$\beta$ ratio is also consistent with being the Case B value.  The Case B ratios provide strong support for the self-shielded broad-line model of Gaskell, Klimek \& Nazarova.  It is proposed that the greatly enhanced Ly$\alpha$/H$\beta$ ratio at very high velocities is a consequence of continuum fluorescence in the Lyman lines (Case C).  Reddenings of AGNs mean that the far-UV luminosity is often underestimated by up to an order of magnitude.  This is a major factor causing the discrepancies between measured accretion disc sizes and the predictions of simple accretion disc theory.  Dust covering fractions for most AGNs are lower than has been estimated.  The total mass in lower mass supermassive black holes must be greater than hitherto estimated.

\end{abstract}


\section{Introduction}

Knowing the reddening of an active galactic nucleus (AGN) is crucial because it affects our ideas of the luminosity of AGNs, the true spectral energy distribution (SED), emission-line ratios, and the detectability of AGNs.  As \citet{Koratkar+Blaes99} commented, even a small amount of reddening has a dramatic effect
at UV wavelengths and changes the shape of the SED.  Without knowledge of the true SED of an AGN we cannot understand AGN accretion discs.  We also cannot use the correct ionizing continuum in modelling emission lines to determine physical conditions and covering factors of the broad-line region (BLR) and torus. Masses of black holes are commonly estimated using the widths of broad lines and the optical luminosity \citep{Dibai77}. An error in the luminosity thus leads to an error in the black hole mass.  Since sizes of the BLR and torus can be found from reverberation mapping and these sizes depend on the square root of the luminosity, distances to AGNs can be found by comparing the inferred luminosities with observed luminosities \citep{Oknyanskij02,Watson+11,Honig14,Yoshii+14}.  However, if AGNs are to be used as cosmological probes, the attenuation needs to be known.  It is thus clear that knowledge of reddening is critical for almost all aspects of studies of the nature of the AGN phenomenon and of the demographics and evolution of supermassive black holes.

Even though the existence of optically-thick dust beyond the accretion disc is an essential part of orientation-dependent schemes to unify thermal AGNs\footnote {In this paper I will only consider ``thermal'' AGNs.  These are AGNs with a relatively high accretion rate whose energy output is dominated by the thermal emission from the accretion disc.  For a discussion of the difference between these AGNs and  ``non-thermal'' AGNs in which mechanical energy output dominates the energy output, see \citet{Antonucci12}.} \citep{Keel80,Antonucci93,Antonucci12}, the reddening of AGNs has been controversial for half a century (see \citealt{MacAlpine85} for a early review).

The reddening question has also been intimately entwined with what has been referred to as ``the hydrogen-line problem.''  Hydrogen lines are the strongest broad lines in the spectra of AGNs.  \citet{Baker+Menzel38} gave general calculations of expected Balmer line intensities under two assumptions:  ``Case A'' where the gas is optically thin in the Lyman lines, and ``Case B'' where the gas is optically thick in the Lyman lines.  In practice, because the Lyman lines are resonance lines and the abundance of hydrogen is high, Case B is the most relevant for most astrophysical situations.  Over the years many calculations have been made of hydrogen emission line intensities under Baker and Menzel Case B conditions and these calculations show good agreement.

Because the Balmer decrement (the intensity ratios of the Balmer lines, in particular H$\alpha$/H$\beta$), is relatively insensitive to temperature \citep{Baker+Menzel38}, the Balmer decrement has long been used as a reddening indicator.  However, \citet{Pottasch60} noted that significant optical depths in the Balmer lines would lead to steeper H$\alpha$/H$\beta$ ratios.  This is because when the lower Lyman lines have very high optical depths (i.e., Case B), H$\alpha$ is scattered but H$\beta$, after being absorbed by the $n = 2$ to $n = 4$ transitions, can be converted to P$\alpha$ and H$\alpha$. This reduces the number of H$\beta$ photons and increases the number of H$\alpha$ photons.  There are similar conversions for the higher-order Balmer lines but because the optical depths decrease sharply as one goes up the Balmer series it is the H$\alpha$/H$\beta$ ratio that is increased first.

In the first study of the intensities of hydrogen lines in an AGN, \citet{Osterbrock+Parker65} found the Balmer decrement of the narrow-line region (NLR) of NGC~1068 to be steeper than Case B.  This immediately raised the question of whether AGNs were reddened, or whether the intrinsic hydrogen line ratios were different from Case B.  Osterbrock \& Parker assumed that there was insufficient internal extinction in NGC~1068 to explain the steep Balmer decrement and instead favoured the observed steep decrement being a consequence of high optical depths in the Balmer lines.

The first spectrophotometry of the BLR of AGNs \citep{Wampler67,Wampler68a} showed that the observed broad-line decrements were also not Case B.  \citet{Wampler67} noted that reddening of $E(B-V) \sim 0.2$ could explain the Balmer decrement of Ton 1542 but also suggested that H$\alpha$ could be collisionally enhanced.  By comparing high- and low-redshift AGNs, \citet{Wampler68a} estimated the Ly$\alpha$/H$\alpha$ ratio as $\approx 2.7$, which would correspond to Ly$\alpha$/H$\beta \approx 10$ rather than the Case B value of $\approx 33$. Starting with the photoionization models of \citet{Davidson72} it was assumed that there was significant collisional enhancement of Ly$\alpha$ and it became standard to assume Ly$\alpha$/H$\beta = 40$.  However, composite spectra of large numbers of AGNs \citep{Chan+Burbidge75,Baldwin77} confirmed that Ly$\alpha$/H$\beta \approx 10$.  The first direct observations of Ly$\alpha$ in a low-redshift AGN \citep{Davidsen+77} and of H$\alpha$ in a high-redshift AGN in the IR \citep{Hyland+78} supported these results.  These results accelerated observational and theoretical work on the hydrogen line problem.  A large amount of effort was put into trying to reproduce steep Balmer decrements and low Ly$\alpha$/H$\alpha$ ratios with optically-thick photoionization models of the BLR (see \citealt{Ferland+Shields85} for a review).

Because the density was clearly high for the BLR (e.g., \citealt{Dibai+Pronik67}) it was natural to assume that collisional and optical depth effects would be important.  However, it was not clear that these effects were important for the much lower density NLR.  \citet{Wampler68b,Wampler71} showed by using [\ion{S}{ii}] $\lambda$4072 and $\lambda$10,320 lines sharing common $^2P_{1/2,\,3/2}$ upper levels (see \citealt{Miller68}) that reddening was substantial and consistent with that implied by the NLR Balmer decrement.  With the advent of modern digital detectors permitting high-quality spectrophotometry, \citet{Koski78} was able to show that, contrary to the early photographic work of \citet{Osterbrock+Parker65}, NLR hydrogen line ratios of Seyfert 2 galaxies are consistent with reddened Case B values.  From the Balmer decrement of NGC~1068 Koski obtained $E(B-V) = 0.52 \pm 0.04$ which was consistent with Wampler's estimate of 0.5 from the [\ion{S}{ii}] lines.  Further comparisons of a variety of independent reddening indicators for additional objects \citep{Gaskell82,Gaskell84,Wysota+Gaskell88} validated the use of the NLR decrement in general as a reddening indicator. Furthermore, photoionization models \citep{Gaskell+Ferland84} showed that the Balmer decrement of the BLR was unlikely to be far from Case B.  There was thus no need to invoke special effects to modify Balmer line ratios for the NLR.\footnote{It is also interesting to note that more recent observations \citep{Phillips06} have failed to confirm any of the earlier alleged cases of high Balmer line optical depths in planetary nebulae.}

Even though there has thus been clear evidence for a long time that the {\em narrow}-line region of AGNs is reddened there has continued to be strong resistance to the idea that the {\em broad}-line region and continuum are reddened.  It is common to correct spectra only for the small amount of reddening along the line of sight in the Milky Way and to assume that any internal reddening in AGNs is negligible (see, for example, \citealt{Bentz+13} and \citealt{Stevans+14}).  There have been several reasons for the belief that internal reddening is negligible.  As \citet{Woltjer68} pointed out, if the lines are reddened, then the continuum is probably reddened too and this makes AGNs an order of magnitude more luminous.  \citet{Wampler68b} also pointed out that de-reddening of the continuum would produce a spectrum that rose towards the ultraviolet, which was considered unlikely prior to the introduction of the thermal accretion-disc model by \citet{Lynden-Bell69}.  \citet{McKee+Petrosian74} gave two strong additional arguments against AGN spectra being reddened.  The first was a lack of the $\lambda$2175 interstellar absorption feature in AGN spectra and the second was the apparent lack of steep downward curvature of the continuum going into the far UV (see also \citealt{Cheng+91}).  The success of sophisticated photoionization models such as {\it Cloudy} \citep{Ferland+98,Ferland+13} in producing steep Balmer decrements under BLR conditions also made reddening of the BLR seem unnecessary for explaining hydrogen line ratios.

Despite the strong arguments against reddening of the BLR and continuum, \citet{Shuder+MacAlpine77,Shuder+MacAlpine79} argued that the solution to the hydrogen
problem was that both the BLR and continuum of AGNs suffer substantial extinction of the order of $E(B-V) \sim 0.5$.  They suggested that such high reddening could be reconciled with the lack of evidence for reddening in the UV if the attenuation curve of dust near an AGN differed significantly from an average Milky Way reddening curve.  \citet{Netzer+Davidson79} concurred that substantial reddening could be the cause of the low observed Ly$\alpha$/H$\beta$ ratios.  \citet{Soifer+81} found that the BLR Ly$\alpha$/H$\alpha$ ratio depended on the optical to UV continuum slope in a manner consistent with a reddening of $E(B-V) \thickapprox 0.3$.

Observational support for substantial BLR and continuum reddenings in AGNs came from the discovery of correlations of continuum colours and Balmer decrements with host-galaxy inclination, \citep{Cheng+83,deZotti+Gaskell85}.  These showed that there was significant {\em optical} reddening of at least $E(B-V) \sim 0.25$  from dust in the planes of the host galaxies \citep{deZotti+Gaskell85}.  Over the next decade a variety of studies pointed to typical AGN reddenings of $E(B-V) \sim 0.3$ (see section 4.2 below) but the continuing lack of $\lambda$2175 absorption and, in particular, the lack of a strong downwards curvature of the spectrum into the far UV continued to point towards a lack of significant internal reddening \citep{Cheng+91}.  The discovery that the reddening curve of dust in the Small Magellanic Cloud (SMC) lacked the $\lambda$2175 absorption feature \citep{Rocca-Volmerange+81,Nandy+82,Prevot+84} showed that it was possible for there to be dust without this feature, but the SMC reddening curve rises even more steeply into the far ultraviolet than Galactic curves and so should produce a strong downward curvature in AGN spectra.

The solution to the reddening dilemma was provided by \citet{Gaskell+04} who showed, as \citet{Shuder+MacAlpine79} had speculated, that the reddening curve of dust close to the inner regions of an AGN is radically different from previously known curves.  Not only did the reddening curve lack the $\lambda$2175 absorption feature, but the curve was essentially flat in the UV and showed no rise into the far UV (see also \citealt{Czerny+04}, \citealt{Gaskell+Benker07},  \citealt{Lyu+14}, and \citealt{Gaskell+16}).  For a typical face-on, radio-selected AGN \citet{Gaskell+04} estimated the reddening to be $E(B-V)$ $\thickapprox 0.30$.

In this paper I first estimate the velocity-integrated, unreddened H$\alpha$/H$\beta$ and Ly$\alpha$/H$\beta$ ratios for the BLR and show that, contrary to what has been generally assumed for the last half century, the ratios are consistent with the Case B values.  I then discuss the implications of this for the reddening of AGNs, the structure of the BLR, and dust covering factors.  I argue that the hydrogen line ratios emitted by the highest velocity gas close to the black hole are Case C.  I propose that the discrepancy in accretion disc sizes found by reverberation mapping and microlensing when compared with classical accretion disc models is mostly explained by reddening.

\section{The Balmer decrement}

The most widely quoted line ratio for AGNs is H$\alpha$/H$\beta$.  It has been widely assumed, for reasons just outlined, that the unreddened ratio is steeper than Case B.  Optically-thick photoionization models also show that the intrinsic ratio varies with physical conditions (see for example, \citealt{Snedden+Gaskell07}). Are these assumptions correct?  To find the unreddened H$\alpha$/H$\beta$ ratio one cannot simply look for the lowest reported H$\alpha$/H$\beta$ ratios since these will be outliers due to measuring errors.  Since the size of the BLR is not much greater than the radius of the accretion disc producing the optical continuum emission (light weeks compared with light days, say), it is reasonable to assume that the lines and the continuum have similar reddenings and therefore to look at the AGNs with the bluest continua.  Measuring errors might cause some extremely blue AGNs might be outliers in the distribution of {\em continuum slopes} but this will not affect the estimation of the H$\alpha$/H$\beta$ ratios.

Contrary to expectations, \citet{Dong+08} have found from a careful study of a sample of blue AGNs (which they considered to have negligible reddening) that the velocity-integrated H$\alpha$/H$\beta$ ratio shows a surprisingly low observed scatter of only 0.05 dex\footnote{\citet{Baron+16} find an observed scatter in H$\alpha$/H$\beta$ for the bluest AGNs they consider that is over three times larger.  Given the similarity of the \citet{Dong+08} and \citet{Baron+16} samples,  this points to relatively larger errors in the fluxes in the \citet{Shen+11} catalogue used by \citet{Baron+16}.}.  This implies an intrinsic scatter of only 0.03 dex after allowance for their estimated observational errors of 0.04 dex.  The small scatter \citet{Dong+08} find strongly supports the use of H$\alpha$/H$\beta$ as a reddening indicator for the BLR.

If we accept that the observed velocity-integrated H$\alpha$/H$\beta$ ratio primarily depends on reddening, it is reasonable to ask to what extent the blue AGN sample of \citet{Dong+08} really is unreddened and whether the intrinsic velocity-integrated H$\alpha$/H$\beta$ ratio is actually {\em less} than the value of 3.06 they propose.  I argue here that this is indeed the case.

\subsection{Object selection}

H$\alpha$ lies within the spectral coverage of the Sloan Digital Sky Survey (SDSS) for $z < 0.35$.  \citet{Dong+08} chose a sample of $\thickapprox 4100$ SDSS DR4 spectra of AGNs within this redshift range and with a median signal-to-noise (S/N) ratio per pixel $\geqslant 10$.  From these they selected the 446 AGNs with spectral index $\alpha_{\nu} > -0.5$ (where $F_{\nu} \propto \nu^{+\alpha_{\nu}}$) $\alpha_{\nu}$ is measured from $\lambda$4030 to $\lambda$5600).  This corresponds to the bluest $\sim 11$\% of the AGNs in the SDSS sample.  For these they present the intensities and many profile parameters of H$\alpha$ and H$\beta$ after careful subtraction of contaminating lines etc.\@ (see \citealt{Dong+08} for details).  Because the AGNs are very blue, host-galaxy starlight was assumed to be negligible.

In order to investigate the extent to which the blue AGN sample of \citet{Dong+08} is reddened, I have constructed a subset of extremely blue AGNs with $\alpha_{\nu} > +0.2$.  This extremely blue sample consists of 27 AGNs or about 0.6\% of SDSS AGNs with $z < 0.35$ and S/N $\geqslant 10$. \citet{Dong+08} noted that their blue AGN sample is of systematically higher luminosity compared to less blue, low-redshift SDSS AGNs.  This could be because reddening decreases with luminosity as suggested by \citet{Gaskell+04}. To reduce the effect of possible luminosity-dependent reddening, and possible dependence of the intrinsic spectral slope on luminosity, a luminosity-matched control sample was constructed from the $\alpha_{\nu} < +0.2$ sample for comparison purposes. This subset was constructed as follows: for each of the bluest AGNs, two control AGNs were taken from the $\alpha_{\nu} < +0.2$ AGNs, the one immediately above the bluest AGN in luminosity and the one immediately below it in luminosity.  \citet{Dong+08} estimate a typical error in $\alpha_{\nu}$ to be $\thickapprox \pm 0.04$.

\subsection{Analysis}

Properties of the $\alpha_{\nu} > +0.2$ AGNs were compared with those of the luminosity-matched subset of $-0.5 < \alpha_{\nu} < +0.2$ AGNs.  Properties compared were the redshift ($z$), the black hole mass, the Eddington ratio, the luminosity of H$\alpha$, the equivalent width of H$\alpha$, the Balmer decrement, the curvature of the optical spectrum, the fluxes in the {\it GALEX} FUV ($\lambda$1528) and NUV ($\lambda$2271) passbands, the UV to optical spectral shape, the radio luminosity, and line profile parameters (line asymmetry, kurtosis, shape, FWHM, and dispersion -- see \citealt{Dong+08} for definitions).

\citet{Dong+08} fit a single log-normal function to the distribution of Balmer decrements. However, given that they made an arbitrary continuum colour cutoff at $\alpha_{\nu} = -0.5$ and since the observed (uncorrected) Balmer decrement will be correlated with the observed continuum colour when there is reddening, the relative lack of steep H$\alpha$/H$\beta$ ratios is simply a consequence of the colour cutoff.  If redder AGNs ($\alpha_{\nu} < -0.5$) had been included, the distribution would extend to much higher H$\alpha$/H$\beta$ ratios.  Such ratios are well known in other samples (for example, \citealt{Dong+05} present cases of AGNs with extremely large H$\alpha$/H$\beta$ ratios).  The value of the mean H$\alpha$/H$\beta$ ratio found by \citet{Dong+08} is thus affected by the continuum colour cutoff of the sample.

In order to estimate the unreddened H$\alpha$/H$\beta$ ratio let us make three assumptions: (a) the velocity-integrated H$\alpha$/H$\beta$ has a precise value, (b) steeper Balmer decrements are only due to reddening, and (c) there is a flat distribution of $E(B-V)$ for AGNs.  The distribution of reddened  H$\alpha$/H$\beta$ ratios will then be a step function.  The observed distribution will be this step function convolved with scatter due to measurement error and intrinsic spread.  By fitting this to the low H$\alpha$/H$\beta$ side of the distribution, the unreddened ratio and the scatter can be estimated under these idealized assumptions.   The assumption that the distribution of reddening is a step function is a conservative one.  The result does not depend strongly on the assumed distribution of reddenings.  If the distribution falls off with increasing reddening (i.e., there are fewer AGNs with higher reddenings), the analysis gives a somewhat higher intrinsic Balmer decrement.  However, there is a limit to this.  The extreme case is that the distribution of reddenings is a $\delta$-function at zero.  This corresponds to the assumption of \citet{Dong+08} that their blue AGNs are unreddened.  A rising distribution of reddening (unreddened AGNs are rare) gives a lower intrinsic decrement.  The limit of this is where there is a rising distribution with no scatter due to observational errors.  In this limit the distribution of ratios is the just the distribution of reddenings and the intrinsic H$\alpha$/H$\beta$ ratio is the lowest observed value.

\subsection{The intrinsic Balmer decrement}

The distributions of H$\alpha$/H$\beta$ ratios for the entire \cite{Dong+08} sample and for the $\alpha_{\nu} > +0.2$ subsample are shown in Figure 1.  As expected, the bluest AGNs (shown as a histogram at the bottom of Figure 1) have a significantly flatter Balmer decrement.  The geometric mean ratio is H$\alpha$/H$\beta = 2.86 \pm 0.06$.  The corresponding geometric mean ratio for the luminosity-matched subset of the \cite{Dong+08} sample is H$\alpha$/H$\beta = 3.13 \pm 0.05$ (consistent with the mean of 3.06 for the entire \citealt{Dong+08} SDSS sample and the similar, but less certain, value of $3.02 \pm 0.18$ of \citealt{Baron+16}).  The difference between the $\alpha_{\nu} > +0.2$ AGNs and the luminosity-matched $\alpha_{\nu} < +0.2$ AGNs is highly significant ($p = 0.0004$).

% Figure 1
\begin{figure}
 %\vspace{302pt}
 \centering
 \includegraphics[width=8cm]{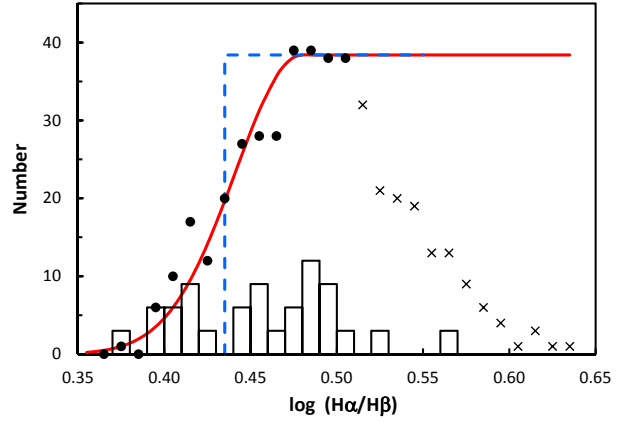}
 \caption{Observed and calculated distributions of BLR Balmer decrements.  The solid circles and crosses show the distribution of observed H$\alpha$/H$\beta$ ratios for the entire blue AGN sample of \citet{Dong+08}.  These correspond to the bluest 11\% of $z < 0.35$ AGNs observed by the SDSS.  The dashed (blue) curve shows a hypothetical distribution assuming that the intrinsic unreddened H$\alpha$/H$\beta$ ratio is 2.72, that all values greater than this are due to reddening, and that there is a flat distribution of reddenings. The smooth (red) curve shows the same distribution if there is scatter of 0.055 dex due to measuring errors etc.  This has been fit to the points shown as solid circles.  The histogram at the bottom shows the distribution of Balmer decrements for the bluest AGNs alone ($\alpha > +0.2$).  This has been multiplied by a factor of four for ease of comparison.}
\end{figure}

Figure 1 also shows the fit  of the hypothetical step-function distribution discussed in Section 2.2 and its convolution with a log-normally distributed scatter.  This is fit to the low H$\alpha$/H$\beta$ side of the observed distribution (the solid black circles).  This gives an unreddened H$\alpha$/H$\beta$ ratio of $2.72 \pm 0.04$, which is slightly lower than the geometric mean for the very blue AGNs. As noted above the intrinsic ratio derived is not sensitive to the assumed form of the distribution of reddenings.  The extreme case of a declining distribution is the \citet{Dong+08} vaue of H$\alpha$/H$\beta = 3.06$.  However, this is greater than the geometric mean ratio of 2.86 for the very bluest AGNs considered here so this is unlikely.  The limit of the assumption that there is a rising distribution and no observational errors gives H$\alpha$/H$\beta = 2.5$.  These limiting cases thus support the intrinsic value being $\thickapprox 2.7$ as given by the assumption of a flat distribution of reddenings.

Comparison of other properties between the bluest AGNs and the luminosity-matched control sample shows no significant differences for redshift, mass, Eddington ratio, ultraviolet spectral shape (as given by the FUV-NUV colour index), line widths or profile shapes.  The only significant differences ($p \leqslant 0.01$) are in the UV luminosities, which are higher for the bluest AGNs, particularly compared with the $i$ band, and in the ratio of 2 $\mu$m flux to $u$-band flux, which is higher for the redder AGNs.  The bluest AGNs also have lower radio power but this is only marginally significant ($p = 0.07$).

\subsection{The Balmer decrement of the most luminous AGNs}

As noted, reddening decreases with luminosity (see \citealt{Gaskell+04}) so the least reddened AGNs should be those with the highest luminosities.  An independent derivation of the bluest H$\alpha$/H$\beta$ ratio was made by looking at the highest luminosity AGNs from the larger SDSS DR7 sample of \cite{Shen+11}.  For the bluest half of AGNs with $\log L_{Bol} > 46$ (2.5\% of the DR7 AGNs with $z < 0.35$) the median H$\alpha$/H$\beta = 2.74 \pm 0.11$ using the line intensity measurements from \citet{Shen+11}.  This is in good agreement with the value derived above from the very blue subsample of measurements of \citet{Dong+08}.

\section{Lyman $\alpha$/H$\beta$}

\citet{Soifer+81} found that Ly$\alpha$/H$\alpha$ was correlated with the optical to UV continuum slope in a manner suggestive of common reddening of the BLR and continuum.  Assuming a Case B Ly$\alpha$/H$\alpha$ ratio and using a Galactic reddening curve this corresponded to a range of reddening of $0.19 < E(B-V) < 0.44$. These results were confirmed with additional observations by \citet{Puetter+81} and \citet{Allen+82}.  \citet{Kollatschny+Fricke83} noted that $0.20 < E(B-V) < 0.25$ could explain the Lyman, Balmer, and Paschen line ratios of PKS 1011-282 and PKS 1217+023.

Figure 2 shows the Ly$\alpha$/H$\beta$ ratio versus continuum slope from the combined data sets of \citet{Wills+93}, \citet{Wills+95}, \citet{Netzer+95}, and \citet{Bechtold+97}.  These data have the advantage of higher signal-to-noise ratio than the early observations because the UV spectra come from the {\it Hubble Space Telescope} rather than from the {\it International Ultraviolet Explorer} for Lyman $\alpha$ observations of low-redshift AGNs, and there were also major improvements in IR detectors for observations of H$\beta$ in the high-redshift AGNs \citep{Bechtold+97}.
The rest-frame UV and optical observations of the samples were also effectively simultaneous.  The Galactic foreground reddening corrections calculated by Wills et al.\@ used the old reddening estimates of \citet{Burstein+Heiles82} which systematically underestimate the Galactic reddening at high latitudes compared with more recent measurements of \citet{Schlafly+Finkbeiner11}.  This small difference -- an average increase in the Ly$\alpha$/H$\beta$ ratio of $\approx 6$\% -- is unimportant here.

% Figure 2
\begin{figure}
 %\vspace{302pt}
 \centering \includegraphics[width=8.5cm]{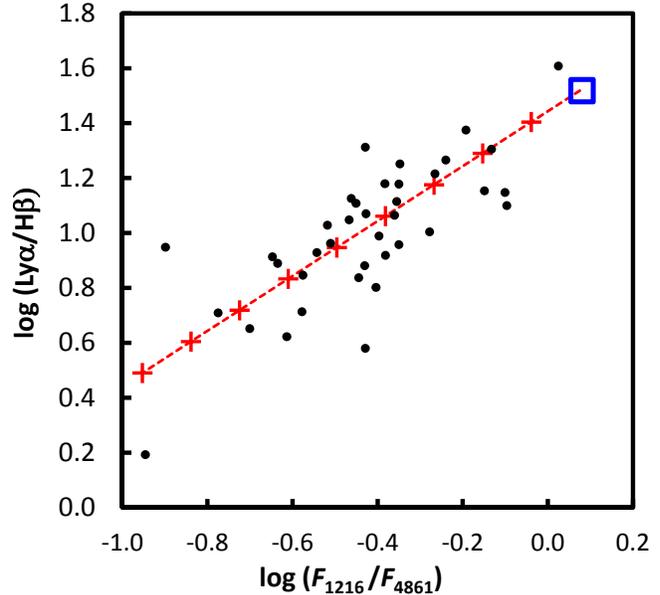}
 \caption{ Ly$\alpha$/H$\beta$ ratios versus the $F_{1216}/F_{4861}$ continua flux ratios (in flux per unit frequency) from the observations of \citet{Netzer+95} and \citet{Bechtold+97}.  The blue square in the upper right corner corresponds to the Case B Lyman$\alpha$/H$\beta = 33$ and $F_{1216}/F_{4861} = 1.19$ (i.e., $F_{\nu} \propto \nu^{+0.24}$) for $E(B-V) = 0.00$.  The red dashed line is the reddening curve.  The crosses indicate steps of $\Delta E(B-V) = 0.10$ given by the \citet{Weingartner+Draine01} reddening curve for a ratio of total to selective  extinction, $R_V \equiv A_V/E(B-V) = 5.5$.  For a solar neighbourhood reddening curve with $R_V = 3.1$ the crosses would be approximately twice as far apart (i.e., the crosses in the diagram would correspond to steps of $\Delta E(B-V) \approx 0.05$). }
\end{figure}

From Figure 2 it can be seen that the observed continuum colors and line ratios of AGNs are readily explained by a Case B intrinsic ratio with different reddenings.  For these samples the median $E(B-V) \sim 0.2 - 0.4$ depending on the form of the extinction curve used.  Note that neither set of observations was corrected for the host galaxy starlight contribution to the continuum flux at $\lambda$4861.  This does not affect the line ratios but the host galaxy starlight plus measuring errors will introduced some scatter into the relationship in Figure 2.  The scatter of the continuum flux ratio about the reddening line in Figure 2 is only $\pm 0.17$ dex (i.e., less than 50\%).  This is consistent with observed object-to-object AGN continuum to starlight ratios in typical spectroscopic apertures (see, for example, Table 12 of \citealt{Bentz+13}).

\section{The reddening of AGNs}

\subsection{The differences in BLR line ratios are due to reddening, not intrinsic properties}

The lack of differences between the extremely blue, low-redshift SDSS AGN sample constructed here, and the luminosity-matched control sample (see section 2.1) indicate that {\em there are  no fundamental differences in the properties of the two samples other than those caused by reddening.}  As noted, the AGNs are indistinguishable by mass, Eddington ratio, or line profile (which depends on the Eddington ratio and the orientation).  The marginally-significant difference in radio power is attributable to the spectroscopy targeting algorithm of the SDSS. Radio-loud AGNs in the SDSS are steeper than radio-quiet ones because the SDSS preferentially targeted point-source matches to {\it FIRST} radio sources {\em without reference to their colours}, whilst radio-quiet AGN candidates could only be selected by having non-stellar colours (see \citealt{Richards+02} for details).

The most interesting similarity between the two samples is the similarity in (FUV - NUV) colour index.  Because the FUV and NUV observations were simultaneous, variability is not a problem. The starlight contribution from stars in the bulge of the host galaxy is also unimportant in the UV.  The mean (FUV - NUV) colour indices for the bluest AGNs and the control sample are 0.27 and 0.29 respectively.  The similarity argues strongly that the very blue optical continuum of the bluest AGNs is {\em not} the result of a harder UV spectrum.

As an additional check the the ratios of 2MASS $K$-band fluxes to SDSS $g$-band fluxes were compared for AGNs from the larger SDSS DR7 sample.  These pass bands correspond roughly to the $i$ and $H$ bands in the rest frames of the AGNs.  For AGNs with H$\alpha$/H$\beta < 2.8$ the mean ratio  $F_K / F_g$ was $2.75 \pm 0.22$ whilst for AGNs with $ 2.75 <$ H$\alpha$/H$\beta < 3.15$ (i.e., close to the unreddened ratio \citealt{Dong+08} suggest), the mean ratio was $3.68 \pm 0.33$.  This greater ratio means that the optical continuum is, on average, 34\% brighter relative to the IR flux in the bluest AGN subset of the \citet{Dong+08} sample.  For a standard Milky Way reddening curve this corresponds to a difference of $E(B-V) \thickapprox 0.10$.  The ratio of optical to IR fluxes thus provides independent support for the unreddened H$\alpha$/H$\beta$ ratio being closer to 2.72 rather than 3.05.

% Details in my file DR7_Ha_MgII see Name_L_E_Balmer

\subsection{Are the reddenings implied by the Balmer decrements reasonable?}

Many studies using independent indicators of reddening point to optically-selected AGNs typically having reddenings of E(B-V) $\thickapprox 0.2 - 0.3$.  These include:

\begin{enumerate}

\item \citet{Ward+87} proposed that AGNs have a single underlying form that is modified by extinction.  From the optical and IR continuum colours they give for 27 AGNs, the median reddening is $E(B-V) = 0.25$ (see their Fig.\@ 4).

\item \citet{Carleton+87} showed that reddenings estimated from broad H$\alpha$/H$\beta$, from the ratio of 6 keV X-rays to H$\alpha$, and from the 0.36\,$\mu$m/1.2\,$\mu$m continuum colour were well correlated.  From their Fig.\@ 6 the median reddenings from the three methods are $E(B-V) = 0.35$ for their sample of 37 AGNs with hard X-ray measurements.

\item \citet{Winkler+92} and \citet{Winkler97} estimated the continuum reddening of 30 AGNs from optical continuum variability.  The median of their estimates is $E(B-V) = 0.22$.

\item \citet{Xie+16} find from a study of the SEDs of a large number of AGNs at intermediate and high redshifts (samples with $0.71 < z < 1.19$ and $1.90 < z < 3.15$ respectively) that the reddening of their lowest luminosity AGNs (luminosities comparable to the median for the $z < 0.35$ AGNs considered here) are in the range $E(B-V) \sim 0.19$ to $E(B-V) > 0.30$ depending on the form of the reddening curve assumed.

\item Radio emission is a powerful tool for estimating the orientations of AGNs (see, for example, \citealt{Marin14}) and the important study of \citet{Baker97} showed that extinction is very high ($E(B-V) \sim$ 1 to 1.5) in AGNs viewed at high inclinations (lobe-dominated sources). A great merit of her study is the selection by optically-thin radio emission, minimizing selection biases with respect to orientation

\end{enumerate}

For all these cases except for \citet{Xie+16} a standard Galactic reddening curve with a ratio of total to selective extinction of $R_V \equiv A_V/E(B-V) = 3.1$ has been assumed.  It is also important to note that the median reddening estimates quoted above are all {\em lower limits} since it has to be assumed that the bluest objects are unreddened.  It is clear that the reddening implied by these various optical, IR, radio, and X-ray studies are all consistent with the reddenings implied by observed BLR Balmer decrements and Ly$\alpha$/H$\beta$ ratios if we assume that the intrinsic ratios are Case B. All the evidence points to the average low-luminosity AGN having a reddening more than an order of magnitude greater than the reddening due to the Milky Way alone since typical foreground reddening at high Galactic latitudes is only a couple of hundredths of a magnitude.

\subsection{The amount of extinction}

The median observed H$\alpha$/H$\beta$ ratio for typical optically-selected, nearby AGNs (e.g., for Markarian Seyferts and NGC objects of \citealt{Osterbrock77}) is $\approx 4$.  This implies a typical reddening for these objects of $E(B-V) \sim 0.30$.  The reddening from Ly$\alpha$/H$\beta$ depends on the shape of the reddening curve.  Adopting the \citet{Gaskell+Benker07} curve gives $E(B-V) \approx 0.28$ mag for the median Ly$\alpha$/H$\beta$ ratio in Figure 2.  For the \citet{Weingartner+Draine01} $R_V = 4$ curve it is 0.25 mag.\@ and for their $R_V = 5.5$ curve it is 0.43 mag.  This range of reddenings is thus similar to the reddening implied by H$\alpha$/H$\beta$ for typical optically-selected, low-redshift AGNs.

These reddenings mean that AGN luminosities are seriously underestimated if internal reddening is ignored.  For H$\alpha$/H$\beta = 4.0$, even if $R_V$ is only 3.1, (the typical Galactic value) there is a whole magnitude of extinction at $\lambda$5100.  If $R_V = 5.5$ the extinction at $\lambda$5100 is 1.4 magnitudes.  This means that luminosities at $\lambda$5100 for low-luminosity AGNs are underestimated by a factor of $\sim 4$.  At $\lambda$1450 the underestimate of the luminosity is a factor of 13.3 -- 6.4 for $R_V =$ 3.1 -- 5.5 respectively.\footnote{This has the opposite dependence on $R_V$ than the optical extinction because reddening curves with a higher $R_V$ are flatter in the UV.}  Since the AGN reddening curve is flat shortwards of $\lambda$1450 this means that the bolometric luminosity of well-studied, low-luminosity AGNs (such as NGC~5548) is often underestimated by an order of magnitude or more.

The largest sample of AGNs used for statistical purposes is now from the SDSS.  These are strongly biased towards blue objects \citep{Richards+02}.  They also include many objects of higher luminosity.  Not surprisingly therefore, the median H$\alpha$/H$\beta$ ratio for AGNs in the SDSS is lower than for the well-studied nearby Seyferts.  For the DR7 AGNs the median ratio is 3.48.  If the unreddened H$\alpha$/H$\beta$ ratio is 2.72, the median reddening of the DR7 sample is $E(B-V) = 0.19$.  This corresponds to a factor of 2 -- 2.5 in the $\lambda$5100 flux and a factor of 3.4 -- 5.6 in the $\lambda$1450 flux.
\footnote{\citet{Baron+16} give the reddening of a typical SDSS DR7 AGN based on observed H$\alpha$/H$\beta$ ratios as only $E(B-V) \thickapprox 0.08$.  This value is lower than what is given here because they assumed the steeper intrinsic Balmer decrement of \citet{Dong+08}.\citet{Baron+16} also argue that the modest extinction they propose arises in the interstellar medium of the host galaxy.  Their argument for this is based interpreting \ion{Na}{I} D and other absorption lines they identify in stacked SDSS spectra as interstellar absorption lines in the host galaxies.  However, consideration of spectra and measurements of line strengths for local K giants and globular clusters (see \citealt{Burstein+84} and \citealt{Faber+85}) points to \ion{Na}{I} D having a predominantly {\em photospheric} origin in stars of the host galaxies.  The \ion{Na}{I} D equivalent widths \citet{Baron+16} give are similar to the equivalent widths of stellar \ion{Na}{I} D in normal galaxies (see \citealt{Trager+98}). The variation in equivalent widths of \ion{Na}{I} D with AGN properties found by \citet{Baron+16} is then due to varying amounts of host-galaxy starlight, not to varying interstellar absorption in the host galaxies.}

\subsection{The reddening curve for AGN dust}

By construction, the mean $\lambda$5100 luminosities of the bluest SDSS AGNs and the control sample considered in section 2.2 are identical.  AGNs are variable and the {\it GALEX} observations were not made at the same time as the SDSS spectroscopy, but nonetheless for the bluest AGNs the (geometric) mean  FUV and NUV fluxes are greater than those of the rest of the \citet{Dong+08} blue AGN sample by a factors of $2.4 \pm 0.8$ and $2.2 \pm 0.6$ respectively.  These numbers are only an upper limit to the relative extinctions (colour excess) between $\lambda$5100 and the {\it GALEX} bands because \citet{Dong+08} did not subtract out the host galaxy starlight contribution to the flux at $\lambda$5100 (i.e., the greater extinction of the control sample produces less of a change in the $\lambda$5100 flux than in the UV fluxes because the $\lambda$5100 flux includes a starlight contribution). For $E(B-V)$ of 0.07, the \citet{Czerny+04} and \citet{Gaskell+Benker07} reddening curves (which are essentially identical after normalization -- see Figure 1 of \citealt{Gaskell+16}) predict that the FUV flux will be 1.3 times greater.  This is consistent with the observed limit.  If AGNs are reddened by an SMC-like reddening curve, as advocated by \citet{Richards+03}, \citet{Hopkins+04} and \citet{Krawczyk+14}, the FUV will be $\thickapprox 2.5$ times greater, which is marginally consistent with the observed limit.

The similarity of the (FUV - NUV) colour index for the two samples is the most important constraint for the reddening curve.  {\em A steep, rising, SMC-like reddening curve is strongly ruled out} because it predicts a 0.75 magnitude greater mean (FUV - NUV) colour for the control sample whereas the actual difference is only $0.02 \pm 0.07$.  The observed difference in UV colors is in good agreement, however, with the predictions of the \citet{Gaskell+04} and \citet{Gaskell+Benker07} reddening curves ($\thickapprox 0.02$ and 0.13 magnitudes respectively) or of the $R_V = 5.5$ curve of \citet{Weingartner+Draine01} (0.10 magnitudes).  An additional major problem for correcting fluxes with a steep SMC-like reddening curve is that for a typical nearby, well-studied Seyfert with H$\alpha$/H$\beta = 4$, an SMC curve predicts an {\em enormous} reduction in the continuum and line flux at Lyman $\alpha$. This is a factor of 170 if the intrinsic H$\alpha$/H$\beta$ ratio is 2.72 as advocated here, or a factor of 35 if the \citet{Dong+08} ratio is used.

% An attenuation curve that is flat in the UV solves other problems.  For example, \citet{Sprayberry+Foltz92} showed that the redder UV continua of low-ionization,  broad-absorption-line quasars (Lo-BALs) could be explained by reddening, but noted that the low reddening they deduced assuming an SMC reddening curve failed to explain the strong IR-to-optical flux ratios.  A flat UV extinction curve removes this difficulty.  The small change in the UV SED now corresponds to a much larger $E(B-V)$ and can produce an order-of-magnitude increase in the IR-to-UV flux ratio as found by \citet{Low+89}.

\section{Intrinsic hydrogen line ratios}

What is remarkable about the unreddened H$\alpha$/H$\beta$ ratios found here is that, rather than being steeper than Case B, as in the vast number of
optically-thick models calculated over the four decades or so, {\em the ratio seems to be identical to a Case B value.}  Similarly, the Ly$\alpha$/H$\beta$ ratio,
rather than being much less than Case B, is also consistent with being Case B.  In contrast to this, models of line ratios with high optical depths in the Balmer lines do {\em not} naturally produce Case B line ratios.  Inspection of Table 3 of \citet{Mathews+80} or Figure 3 of \citet{Snedden+Gaskell07}, for example, shows that a wide range of ratios are produced depending on the physical conditions.

\subsection{Why Case B?}

I believe that Case B line ratios are a natural consequence of the structure of the BLR.  \citet{Gaskell+07}(GKN) argue that the BLR is a flattened ensemble of small clouds above the plane of the accretion disc and that it is optically thick in the equatorial direction.   As a consequence the inner BLR clouds shield the outer BLR clouds from the ionizing radiation from the inner disc and corona.  This produces strong radial stratification in ionization of the BLR in excellent agreement with observations (see GKN and \citealt{Gaskell09}).  Previous BLR models had a focus on resolving the hydrogen line problem by having very high optical depths.  The individual clouds were optically thick and the models only considered radiation escaping from the illuminated front side of the cloud and the unilluminated back side.  This fails to produce the observed ionization stratification (see \citealt{Gaskell09}). These optically-thick cloud models ignore that fact that photons are at least twice as likely to escape {\em sideways} from the cloud.  In contrast to the standard modelling approach, in the GKN model the {\em ensemble} of clouds is optically thick in the Lyman continuum (i.e., the ensemble matter-bounded in the equatorial direction) but {\em each individual cloud is optically thin.}  Because type 1 AGNs are always seen close to face on, the radiation we see comes directly out from the clouds and hence the path of escape has a very low optical depth.  This is illustrated in the cartoon in Figure 3.

% Figure 3
\begin{figure}
 %\vspace{202pt}
 \centering \includegraphics[width=8.3cm,angle=0]{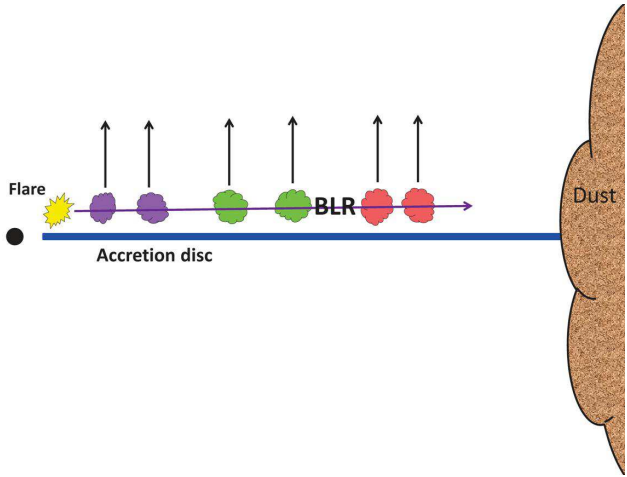}
 \caption{{\bf A.} Cartoon of the BLR model proposed by \citet{Gaskell+07}. The BLR has a flattened distribution above the accretion disc.  Most of the ionizing radiation is generated in the innermost regions and must pass through the BLR clouds as indicated by the horizontal arrow.  This produces the observed strong radial ionization dependence of the BLR.  The individual clouds are optically thin in the Lyman continuum but the optical depth of the ensemble is high in the equatorial direction (i.e., the whole ensemble is matter bounded in the radial direction parallel to the disc.)  The line emission from the clouds is emitted in the vertical direction as indicated by the short arrows.}
\end{figure}

We can thus see that the GKN model {\em requires} Case B line ratios because of the ease of escape of line photons from individual clouds.  The evidence for Case B hydrogen line ratios is thus strong support for the GKN model.

\subsection{The velocity dependence of the Balmer decrement}.

\citet{Shuder82} discovered that the BLR Balmer decrement was significantly flatter in the high-velocity wings (see also \citealt{Crenshaw86} and \citealt{Snedden+Gaskell07}). Given that the decrement for the velocity-integrated profile is well within the range of values for Case B, this means that the H$\alpha$/H$\beta$ ratio in the high-velocity wings (only a small part of the total line flux) is {\em less} than the standard Case B value.  This has a very natural explanation: the Case B Balmer decrement becomes flatter at higher densities.  Figure 4 shows the theoretical Case B H$\alpha$/H$\beta$ ratios as a function of density, $n_{\mathrm{H}}$.  These have been calculated using {\it Cloudy 13.03} \citep{Ferland+13}.  The hydrogen atom of {\it Cloudy} is described in \citet{Ferguson+Ferland97}. For $n_{\mathrm{H}} \lesssim 10^8$ cm$^{-3}$ the ratios agree well with \citet{Storey+Hummer95}.  However, the latter did not include collisional excitation from the ground or first excited states so, for $n_{\mathrm{H}} \gtrsim 10^8$ cm$^{-3}$, the calculations by {\it Cloudy} are to be preferred.  These show that the Balmer decrement becomes much flatter at high densities.  If the higher velocity BLR gas closer to the centre of the accretion disc has a higher density it will therefore have a flatter Balmer decrement as is observed.  The higher density is the cause of the much higher \ion{He}{ii}/H$\alpha$ ratio at high velocities as well (see \citealt{Gaskell+Rojas-Lobos14}.)  The observed high \ion{He}{ii}/H$\alpha$ ratio also requires the high-velocity emitting clouds not to have high optical depths.

% Figure 4
\begin{figure}
 %\vspace{202pt}
 \centering \includegraphics[width=8cm,angle=0]{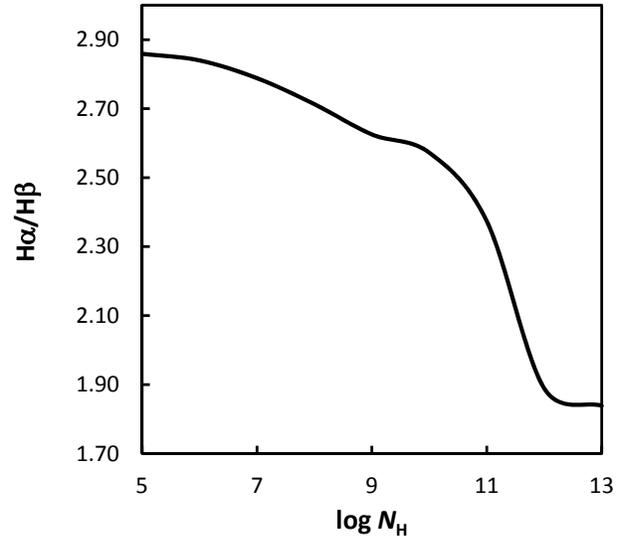}
 \caption{The Case B H$\alpha$/H$\beta$ ratio as a function of hydrogen density, $n_{\mathrm{H}}$, for a temperature of $10^{4}$ K.}
\end{figure}

\subsection{The velocity dependence of Ly$\alpha$/H$\beta$}.

\citet{Zheng92} found that the Ly$\alpha$/H$\beta$ ratio shows a strong dependence on velocity and that the ratio is higher in the high-velocity wings of the lines.  \citet{Snedden+Gaskell07} confirmed this and showed from {\it Cloudy} modelling that there is a strong increase in Ly$\alpha$/H$\beta$ for conditions expected close to the inner regions of an AGN.  Both observationally and in {\it Cloudy} calculations this increase of Ly$\alpha$/H$\beta$ in the high-velocity wings is a much stronger effect than the corresponding decrease in H$\alpha$/H$\beta$.  The observed increase of Ly$\alpha$/H$\beta$ can be a factor of $> 5$ (see Figure 1 of \citealt{Zheng92}).

The Case B calculations of \citet{Storey+Hummer95} do not reproduce this behaviour.  This is because they did not consider continuum fluorescence in the Lyman lines.  The case where this is important has been called Case C \citep{Baker+38,Chamberlain53,Ferland99}.  Unlike Case A and Case B, Case C does not give definite predicted ratios depending on density and temperature.  Instead, for a given continuum shape, the line ratios are a strong function of the optical depths in the Lyman lines (see \citealt{Ferland99}).  For low optical depths the Ly$\alpha$/H$\beta$ can be as great as $150 - 200$ or more.  This is more than fives times the Case B ratio and thus able to explain the highest ratios \citet{Zheng92} found.  There are two questions that must be answered however.  The first is, why does the ratio increase for gas closer to the center of the AGN? And the second is, since low Ly$\alpha$ optical depths are needed, can one get the observed intensities?

The sizes of BLR clouds and the factors governing the sizes are not known {\it a priori}.  If, for the sake of argument, we assume that a typical BLR cloud has a density of $10^{10}$ cm$^{-3}$ and that these clouds are located a few light days from the center of an AGN with a luminosity of $10^{42.5}$ ergs s$^{-1}$ (typical numbers of a low-luminosity AGN), then the solid curves in Figure 5 give the predicted Ly$\alpha$/H$\beta$ ratio as a function of cloud size for different distances from the centre.  The ionizing continuum adopted is the standard \citet{Mathews+Ferland87} AGN continuum.

% Figure 5
\begin{figure}
 %\vspace{150pt}
 \centering \includegraphics[width=8.3cm,angle=0]{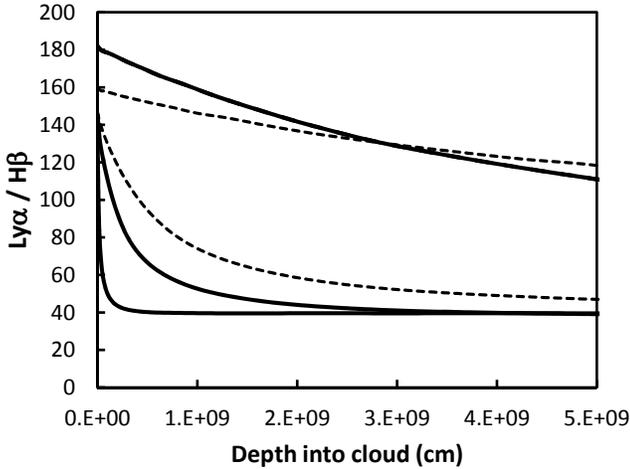}
 \caption{Predicted Ly$\alpha$/H$\beta$ ratios for BLR clouds of density $n_{\mathrm{H}} = 10^{10}$ cm$^{-3}$ but different sizes (given as the depth into the cloud from the surface facing the ionizing radiation).  Results are shown for three distances from the central source of ionizing radiation.  From bottom to top the three solid curves correspond to distances of 10, 3 and 1 light day respectively from the center of an AGN with a luminosity of $10^{42.5}$ ergs s$^{-1}$.  The lower dashed curve is at a distance of 10 light days, but with an internal turbulence of 600 km s$^{-1}$.   The upper dashed curve is for the same location but with turbulence of 10,000 km s$^{-1}$.  The ionizing continuum is the  standard \citet{Mathews+Ferland87} AGN continuum.}
\end{figure}

 As can be seen, the Ly$\alpha$/H$\beta$ ratio is very high for low optical depths (Case C) but then tends to the Case B value for higher optical depths.  The transition from Case C to Case B is after Ly$\alpha$ has become optically thick.  The conditions in Figure 5 span typical BLR conditions.  Reverberation mapping shows that the size of the BLR varies by an order of magnitude from the more distant low-ionization lines such as \ion{Mg}{ii} to high-ionization lines such as \ion{He}{ii} about 10 times closer in (see Figure 5 of \citealt{Gaskell09}).  If the clouds are similar at all radii, the clouds closer to the centre will have a substantially larger zone where Ly$\alpha$ is of low optical depth and the Ly$\alpha$/H$\beta$ ratio is high.

If BLR motions are virialized, as seems to be the case (see \citealt{Gaskell+Goosmann13,Gaskell+Goosmann16}), then a velocity of three times the half width at half maximum (HWHM) of H$\beta$ will correspond to a radius $3^2 \thickapprox 10$ times smaller, i.e., the range of distances considered in Figure 5.  The Ly$\alpha$/H$\beta$ ratio can then be enhanced to several times the Case B value and thus be in agreement with the observations (see Figure 1 of \citealt{Zheng92}).

Despite the success of Case C in explaining the high Ly$\alpha$/H$\beta$ ratios in the high-velocity wings of the line, there is a major problem: the column densities in Figure 5 are very low for a typical BLR distance (the bottom curve in Figure 5).  Even though the ratio of Ly$\alpha$/H$\beta$ is high, the amount of Ly$\alpha$ produced is less than 1\% of the amount produced by radiation-bounded clouds.  A solution to this is to have a high {\it continuum} optical depth through the cloud system (as in the GKN model) while maintaining a low Ly$\alpha$ optical depth in each individual cloud.  This is readily accomplished if there is a velocity difference between the clouds \citep{Gordon+81}.  A cloud behind another cloud can still have strong continuum fluorescence if the Doppler shift causes a small shift in frequency of Ly$\alpha$.  In this manner there is no difficulty in getting Ly$\alpha$ as strong as needed.  This is similar to having individual clouds with a high macroscopic turbulence \citep{Gordon+81}. Figure 5 shows the the effect of turbulence on the Ly$\alpha$/H$\beta$ ratio.  The turbulent BLR calculations of \citet{Bottorff+00} show a similar enhancement in the Ly$\alpha$/H$\beta$ ratio, but curiously, they did not comment on this.  As \citet{Bottorff+00} note, a turbulence of $10^{3}$ km s$^{-1}$ does not violate observations.  By the virial theorem, turbulent velocities will be greater closer to the black hole.  {\em This will be another reason why Case C hydrogen line ratios are much more likely close to the black hole.}  The highest dashed line Figure 5 shows about the maximum turbulence that will be seen in the inner BLR.  Much greater turbulence would give line widths greater than observed.

\section{Dust geometry}

\subsection{The effect of extinction on calculations of dust covering factors from energy balance}

Radiation that is absorbed will be re-radiated in the IR.  The covering factor, $f$, of the dust can be estimated from the ratio of IR luminosity, $L_{IR}$, to bolometric luminosity, $L_{bol}$ (\citealt{Granato+Danese94}; for recent discussion and references to earlier work, see \citealt{Stalevski+16}).  Although $f$ is usually been thought of as a macroscopic covering factor with 100\% absorption of UV radiation within some angle of the equatorial plane and 0\% absorption in the polar directions, $f$ can equally well be thought of as a microscopic covering factor and thus including directions in which the optical depth is not large.   For example, $f = 0.5$ could correspond to 2$\pi$ steradian coverage by 100\% absorption of UV radiation or 4$\pi$ steradian coverage with a 50\% chance of UV photons escaping.  Most of the radiative energy output of AGNs is in the far UV (see, for example, Figure 2 of \citealt{Gaskell08}) but $f$ is calculated either by taking the ratio of {\em optical} luminosity, $L_{opt}$, to  $L_{IR}$, and assuming a bolometric correction to get $L_{bol}$ from $L_{opt}$ (see, for example, \citealt{Maiolino+07}), or by extrapolating to short wavelengths to estimate $L_{bol}$ (e.g., \citealt{Treister+08}).

\citet{Gallagher+07} found $L_{IR}/L_{opt}$ to be increasing with decreasing $L_{opt}$. \citet{Maiolino+07} and \citet{Treister+08} interpreted this non-linear relationship between $L_{opt}$ and $L_{IR}$ as a luminosity-dependence of the covering factor.  However, \citet{Gallagher+07} had showed that if one takes the AGNs with the bluest optical spectra then there is a {\em linear} relationship between $L_{opt}$ and $L_{IR}$.  Since the redder AGNs show a systematically higher $L_{IR}/L_{opt}$ ratio regardless of luminosity, Gallagher et al.\@ interpreted the apparent dependence of $L_{IR}/L_{opt}$ on $L_{IR}$ as luminosity-dependent extinction lowering $L_{opt}$.  \citet{Gaskell+04} had similarly argued that the non-linear relationship between X-ray luminosity and optical luminosity was also due to luminosity-dependent extinction.  Extinction causes the same problem of underestimating $L_{bol}$ when it is estimated by extrapolating to the UV or using {\it GALEX} observations (e.g., as in \citealt{Treister+08}).

\subsection{Eliminating covering factors $>$ 100\%}

The reddenings advocated here have only a small effect on the highest luminosity AGNs, but a substantial effect on low-luminosity AGNs.  This affects calculations of dust covering factors from energy-balance considerations.  From the analysis of \citet{Maiolino+07} the median covering factor for AGNs with $\lambda L_{5100} < 10^{44}$ ergs s$^{-1}$ is 100\%. \citet{Treister+08} get a similarly high covering factor for these AGNs.  A 100\% {\em average} covering factor is physically impossible because any individual AGN must have $f \leqslant 1$ and there is a wide spread in $L_{IR}/L_{opt}$ which will give a wide spread in $f$ among these AGNs.  The one-standard-deviation dispersion in $f$ is $\pm 0.86$ for these lower luminosity Maiolino et al.\@ AGNs and as a result many of them have $f$ much greater than 100\%. Raising $\lambda L_{5100}$ for the lowest luminosity AGNs by a factor of several, as advocated here, removes this problem.

\subsection{Reconciling energetics calculations of covering factors with the fraction of type-2 AGNs}

Another advantage of the extinction corrections proposed here is that they remove the discrepancy between energetics estimates of the dust covering fraction and estimates from the relative fraction of type-2 AGNs at low luminosities. As noted in 6.2.1, if no extinction corrections are applied to the energetics calculations (as was the case, for example, in the estimates of \citet{Maiolino+07} and \citet{Treister+08}) much higher covering factors are found for the lowest luminosity AGNs.  In contrast to this, radio, IR, and volume-limited samples all agree in showing no change in the fraction of type-2 AGNs (about 50\%) with luminosity \citep{Lawrence+Elvis10}.  Allowance for extinction brings the estimates into agreement.

\subsection{No low-redshift ``hot-dust-poor" AGNs}

The bluest AGN subset of the \citet{Dong+08} sample considered in sections 2.1 and 2.2 is expected to have almost zero reddening.  Therefore their ratio of IR to optical flux (see section 6.1) should be close to the intrinsic ratio.  On the basis of optical and near-IR colours \citet{Hao+10} defined a class of what they call ``hot-dust-poor"(HDP) AGNs.  These are about 8 - 10 \% of SDSS AGNs \citep{Hao+11}. They considered most non-HDP AGNs to be essentially unreddened.  However, inspection of the optical and optical--IR spectral indices in Figure 1 of \citet{Hao+10} shows that if their non-HDP AGNs are reddened then their HDP AGNs are consistent with simply being unreddened AGNs.  Compared with the HDP AGNs, typical non-HDP AGNs have a reddening of $E(B-V) \thickapprox 0.2$ as advocated here.  Likewise, Figure 10 of \citet{Hao+11} shows that the $L_{3.5 \mu m}/L_{\lambda 5100}$ ratio is a factor of two less for the HDP AGNs than for SDSS AGNs.  This is again in excellent agreement with the mean extinction being proposed here.

It should be noted, however, that two of the three $z \sim 6$ AGNs identified by \citet{Jiang+10} as lacking near-IR dust emission have a $L_{3.5 \mu m}/L_{\lambda 5100}$ ratio {\em much} lower than the HDP AGNs of \citet{Hao+10,Hao+11} (see Figure 10 of \citealt{Hao+11}).  These very high redshift AGNs do indeed seem to be genuinely lacking hot dust.

\subsection{AGNs with low reddening are very rare}

It is interesting to note that even the bluest 10\% of the low-redshift AGNs targeted by the SDSS (i.e., the \citealt{Dong+08} sample) still have significant reddening.  If the differences in $\alpha_{\nu}$ and H$\alpha$/H$\beta$ are due to reddening this corresponds to the subset of the bluest AGNs (0.6\% of the SDSS DR4 AGNs) having a reddening that is lower by $E(B-V) \thickapprox 0.08$ than the control sample.  Note that we cannot be sure that even this bluest 0.6\% is unreddened.  While $E(B-V) \thickapprox 0.08$ is a modest visual extinction of 0.44 magnitudes (a factor of 50\%) for an $R_V = 5.5$ reddening curve, at $\lambda$1450 the flux has been underestimated by a factor of 1.8.

The bluest 0.6\% of the low-redshift SDSS DR4 AGNs are really a {\em much} smaller fraction of the total AGN population than 0.6\% for a couple of reasons. Firstly, because of the steepness of AGN luminosity functions, if low-luminosity AGNs are reddened by $E(B-V) \approx 0.25$ this will have increased the fraction of the bluest AGNs in a flux-limited sample by a factor of $\approx 6$.  Secondly, colour-based selection will further favour blue objects.  Infra-red selection finds large numbers of reddened AGNs. \citet{Glikman+07} estimate that these heavily reddened AGNs ($E(B-V) \thickapprox 1$) make up 25\% to 60\% of the underlying type-1 AGN population.

We have to remember in addition that more than half of AGNs have the BLR and accretion disc hidden from our direct line of sight (i.e., they are type-2 AGNs or are totally obscured).  For example, \citet{Buchner+15} find that AGNs with $N_H > 10^{22}$ cm$^{-2}$ make up about about $77$\% of the number density and luminosity density of the accreting supermassive black hole population with $L_X > 10^{43}$ erg/s, averaged over cosmic time.

An important thing we can deduce from this extreme rareness of thermal AGNs with low reddening is that {\it the presence of dust above the accretion disc and BLR is a ubiquitous part of the thermal AGN phenomenon.}  This should not be a surprise since dust surrounding an AGN has long been recognized to be an essential part of thermal AGNs \citep{Keel80}.

The dust in AGNs, which has a reddening curve different from typical interstellar dust \citep{Gaskell+04}, has been expelled from the equatorial plane and it is quite likely that this dusty outflow {\em is} what is commonly called the torus (e.g., as proposed by \citealt{Konigl+Kartje94}).  This scenario is supported by mid-IR interferometric observations which show that the dust is elongated on parsec scales in the polar direction \citep{Tristram+12,Honig+12,Honig+13} and that this dust accounts for the {\em majority} of the mid-IR emission (see discussion in \citealt{Honig+13}).

\subsection{Dust scattering cones}

Radiation is also removed from our line of sight by scattering as well as absorption. The albedo of dust grains from the near IR to the near UV is about 0.6 and at short wavelengths the scattering is strongly in the forwards direction \citep{Weingartner+Draine01}.  When the dust is optically thin along our line of sight this scattering will produce small scattering halos around AGNs.  Such halos have been observed. In fact they dominate the light in many AGNs.  The {\it HST} imaging of \citet{Malkan+98} shows that 44\% of type-1 AGNs are what they call ``RS1" (= resolved Seyfert 1) AGNs which show no nuclear point source in {\it HST}.  This is again support for internal dust along the line of sight to type-1 AGNs. \citet{Malkan+98} point out that RS1 AGNs have steeper Balmer decrements and hence greater extinction than non-RS1 AGNs.  In the case of the BAL quasar H1413 + 117 microlensing gives an independent indication of an extended continuum source due to scattering \citep{Sluse+15}.

For type-2 AGNs the half-opening angle of the quasi-conical, extended part of the scattering region can readily be found.  From {\it HST} observations of extended dust scattering cones in 20 luminous type-2 AGNs \citet{Obied+16} find a median half-opening angle, $\theta$, of only $27^{\circ}$ with a standard deviation of only $\pm 9^{\circ}$.  They point out that the half-opening angle should give a fraction of type-1 AGNs of $1 - \cos \theta \thickapprox 10$\% rather than the $\thickapprox 50$\% observed.  They calculate that observational bias only raises the predicted fraction to 13\%. \citet{Obied+16} also point out that the predicted fraction is a lot less than the $(1 - f)$ predicted from energetics.  They suggest that the solution to this discrepancy is that rather than obscuration being ``all or nothing" (0\% for type 1 AGNs and 100\% for type 2 AGNs), many type-1 AGNs are seen through partial obscuration.  This is equivalent to the picture being advocated in this paper. This is illustrated in Figure 6.

% Figure 6
\begin{figure}
 %\vspace{150pt}
 \centering \includegraphics[width=8.3cm,angle=0]{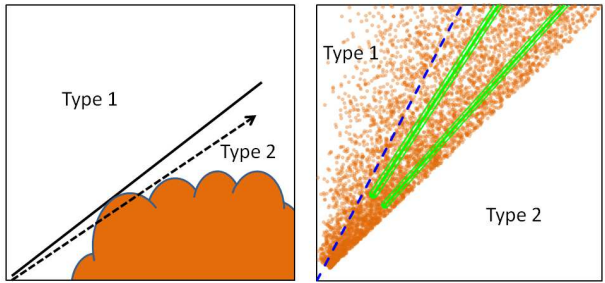}
 \caption{Cartoons illustrating dust in a traditional orbiting ``torus" (left panel) and the outflowing conical distribution proposed here (right panel).  The black hole is at the bottom left corners and the rotation axis is in the $y$ direction.  In the left panel an AGN will appear to be an unreddened type 1 when viewed from above the solid line.  The dotted line illustrated how an AGN could be reddened in this picture.  The dashed blue line in the right panel shows the mean opening angle of scattering cones found by \citet{Obied+16}.  The green lines indicate the median internal and external half opening angles found by \citet{Fischer+13} for bi-conical outflowing NLRs. In the cartoon on the right there is at least some reddening of essentially all type-1 AGNs.  The lower limit to the dust cone in the right-hand panel corresponds to a ratio of Seyfert 2s to Seyfert 1s of 0.7.  Cool molecular gas in the equatorial plane is not shown.}
\end{figure}

\subsection{Location of the dust causing the extinction}

The amount of reddening found here is substantially greater than the reddening of the nuclei of normal early type galaxies.  This suggests that the dust is associated with the active nucleus.  Since the size scale of the NLR is known approximately in AGNs, the relative reddenings of the NLR and the BLR are an important clue to the location of the dust.  \citet{deZotti+Gaskell85} found that the Balmer decrements of the BLR were steeper than those of the NLR (see their Figure 2).  More recent data confirm this and the NLR reddening seems to be an approximate lower limit to the BLR reddening (see Figure 8 of \citealt{Dong+05} and \citealt{Heard+Gaskell16}).  \citet{Heard+Gaskell16} also show that both the equivalent width of [\ion{O}{iii}] $\lambda$5007 and the [\ion{O}{iii}] $\lambda$5007/H$\beta$ ratio increase with the BLR Balmer decrement in a manner which is consistent with the dust being located between the BLR and NLR. There therefore seems to be foreground dust which reddens both the NLR and the inner regions of AGNs, but when there is heavy reddening of the continuum and BLR it is caused by dust located between the BLR and the NLR (or in the innermost part of the NLR).

\section{A model for the creation of region of absorbing dust in AGNs}

A lack of depletions of refractory elements shows that the inner, high-ionization BLR gas is dust free \citep{Gaskell+81}.  Dust can form in the outermost, low-ionization BLR, however \citep{Shields+10}.  The low-ionization BLR is in an outer ring (see \citealt{Gaskell09}).  Most AGNs are within an order of magnitude or so of the Eddington limit.  The standard Eddington limit is calculated for electron scattering.  However, dust grains have a much higher cross section than electrons (see, for example, \citealt{Fabian+08}), so when dust forms it will readily be blown away from the AGN by radiation pressure.  Radiation pressure alone will thus create a dusty, bi-conical outflow starting at the outer edge of the BLR.  Since the grains will be charged they will drag gas with them.   This gas will be seen as the NLR.

The NLR is known to be a relatively thin, bi-conical outflow.  Spatially-resolved kinematic data give the geometry and velocity of this outflow.  \citet{Fischer+13} give half-opening angles of the inner and outer walls of this outflow.  The median half-opening angle is $38^{\circ}$ and the angular thickness of walls of the cone is $\pm 6^{\circ}$.  The angles of the inner and outer walls are shown in the right-hand panel of Figure 6.  It can be seen that these angles are broadly consistent with the half-opening angle predicted by the type-1/type-2 ratio and the half-opening angle of scattering cones (see above).  In real AGNs the dust will have a clumpier distribution than shown in Figure 6.  The maximum outflow velocity is of the order of 800 km s$^{-1}$ \citep{Fischer+13}

Close to the centre of the AGN the dust temperature will be close to the sublimation temperature.  The dust temperature will fall off as the flow moves away from the centre. For type-1 AGNs we observe the side of the dust outflow coming towards us since the far side will be heavily attenuated by the cold, dense molecular gas in the equatorial plane. \citet{Oknyansky+15} find that the IR reverberation lags of the hot dust in response to continuum variability vary little with wavelength and show how this is a natural consequence of the dust being in a bi-conical outflow since the cones will be roughly parallel to the iso-delay surfaces of reverberation mapping (see their Figures 2 and 3).

\section{Other implications of reddening of AGNs}

\subsection{The size of AGN accretion discs}

As was shown by \citep{Lynden-Bell69}, a classical accretion disc has a $T \propto R^{-3/4}$ radial gradient in surface temperature.  By Wien's law the effective emission radius for any wavelength therefore varies as $\lambda^{4/3}$.  Lynden-Bell also showed that such a disc gives an $F_{nu} \propto \nu^{+1/3}$ spectrum.  When the disc reprocesses variable short wavelength radiation from the innermost regions of the AGN the difference in radii producing different wavelengths can be found from the time lags between the responses of the different wavelength regions \citep{Wanders+97,Collier+98} and hence the radius producing a given wavelength can be found.  From a study of wavelength-dependent lags of the optical continua of 14 AGNs \citet{Sergeev+05} discovered that the observed sizes of the discs were systematically larger than predicted by the luminosities of the AGNs (see their Figure 2).  They pointed out that agreement would be better if the absolute magnitudes of the AGNs were about a magnitude brighter.  A detailed re-analysis and interpretation of the \citet{Sergeev+05} observations is given by \citet{Chelouche13}.  The size discrepancy is also confirmed in the recent reverberation mapping study of NGC~5548 by \citet{Edelson+15} and \citet{Fausnaugh+16} and by an analysis of Pan-STARRs light curves of a large number of AGNs \citep{Jiang+16}.

Completely independent confirmation of this discovery has been provided by gravitational microlensing studies.  Microlensing fluctuations are sensitive to the half-light radius of the region responsible for emission at the wavelength of observation \citep{Mortonson+05}.  \citet{Pooley+06} found that microlensing of PG 1115+080 implied a substantially larger size than would be expected for a standard accretion disc.  \citet{Pooley+07} found that this was generally true for microlensed AGNs.  \citet{Morgan+10} and \citet{Jimenez-Vicente+12}, looking at a large number of cases of microlensing, find this size discrepancy to be a factor of four or five.

The median H$\alpha$/H$\beta$ ratio for the 14 AGNs for which \citet{Sergeev+05} estimated wavelength-dependent delays is 4.04.  If we take the unreddened broad-line Balmer decrement to be 2.72, as advocated here and adopt a standard Milky Way extinction curve, this gives a median reddening of $E(B-V)$ = 0.36.
This means that the AGNs of \citet{Sergeev+05} are indeed a magnitude brighter at $\lambda$5100 as they speculated.  However, for dust with a flat reddening curve, $R_V$ will be larger, perhaps as large as $\thickapprox 5.5$ (see \citealt{Gaskell+04}).  In this case the AGNs are $\thickapprox 1.5$ magnitudes brighter at $\lambda$5100.  This is a factor of four in optical luminosity and a factor of $\thickapprox 10$ in the UV.  Since the size of the accretion disc scales as the square root of the luminosity, this naturally explains the larger disc sizes found by \citet{Sergeev+05} and by the more recent microlensing studies. The radius of the region in a disc emitting most of the radiation at a given wavelength also depends on the radial temperature gradient in the disc \citep{Rauch+Blandford91}.  This temperature gradient is flatter than the $T \propto R^{-3/4}$ gradient of a classical accretion disc (see \citealt{Gaskell08}). This causes the region to be somewhat larger than predicted in the classical case but this effect is not enough in itself to explain the observed large sizes  \citep{Hong+16}.

\subsection{Black hole masses and the So{\l}tan argument}

As discussed in 4.3, for a typical SDSS AGN with an internal reddening $E(B-V) \thickapprox 0.2$ the intrinsic flux at $\lambda$1450 is 4 or 5 times greater than observed.  If the attenuation curve to shorter wavelengths is flat, as found by \citet{Gaskell+04}, then the bolometric luminosities are also 4 or 5 times greater.  If the curve is rising to shorter wavelengths (something {\em not} favoured by the observations considered here), the corrections to the bolometric luminosities would be much greater.  An increase in the bolometric luminosity means that the total mass of black holes in the universe estimated by the So{\l}tan argument \citep{Soltan82} has to be increased.  Recent work on the stellar dynamical measurement of black hole masses to which AGN black hole masses are scaled, has also been leading to an upward revision of the black hole mass scale by a factor of three or so for various reasons \citep{Kormendy+Ho13}.  If the total bolometric luminosity of AGNs and the black hole masses are both increased by comparable amounts then there is no change in the mean radiative efficiency.

\subsection{The intergalactic medium ``photon underproduction crisis''}

The number density of AGNs and the typical SED are important factors influencing the intergalactic radiation field (e.g., \citealt{Haardt+Madau96,Haardt+Madau12,Madau+Haardt15}). Hard radiation from AGNs could dominate the reionization of the universe with little need for a contribution from normal star-forming galaxies \citep{Madau+Haardt15}. \citet{Kollmeier+14} have shown that comparison of the most recent statistics for the low-redshift Ly$\alpha$ forest with simulations of the intergalactic medium (IGM) require the metagalactic photoionization rate to be a factor of five higher than had been predicted.  They call this the ``photon underproduction crisis''.  If our ideas about the low-redshift IGM are not to be abandoned then the flux of ionizing photons at low redshifts must be increased.  The reddenings argued for here offer a natural solution to this problem without the need to invoke an undetected population of AGNs, or to invoke exotic new sources of photons (e.g., decaying or annihilating dark-matter particles).  Because the extra photons are coming from a change in the far UV SED, there is no difficulty in satisfying the constraints of the X-ray background (i.e., the X-ray SED of AGNs is unchanged).  By considering the \ion{He}{II} proximity effect in the intergalactic medium, \citet{Zheng+15} also conclude that the 1 -- 4 Rydberg radiation field from AGNs is considerably harder than normally assumed.

\section{Conclusions}

Conclusions are as follows:

\begin{enumerate}

\item Even the bluest 10\% of SDSS AGNs still have significant reddening.  The unreddened, velocity-averaged Balmer decrement of the broad-line region of AGNs is probably a Case B value of H$\alpha$/H$\beta \thickapprox 2.72$.

\item The correlation between the Ly$\alpha$/H$\beta$ ratio and optical to UV spectral index supports the velocity-averaged Ly$\alpha$/H$\beta$ ratio also being Case B.

\item The UV colours of the bluest SDSS AGNs considered here argue strongly against the extinction curve of AGN dust being SMC-like.  Instead, the observations are consistent with the flat AGN reddening curve of \citet{Gaskell+04}.  These observations also support the idea that the intrinsic UV/optical/IR colours of AGNs are similar.

\item Case B hydrogen ratios are a natural consequence of the self-shielding BLR model of \citet{Gaskell+07} where the ensemble of BLR clouds is optically thick parallel to the accretion disc but individual BLR clouds are optically thin and line photons readily escape perpendicular to the disc.

\item The flatter Balmer decrement in the high-velocity wings of the lines is a consequence of higher densities closer to the black hole.

\item The greatly enhanced Ly$\alpha$/H$\beta$ ratio in the high-velocity wings is a consequence of Lyman line fluorescence (i.e., Case C conditions) in the optically-thin gas closest to the black hole.

\item The ensemble of clouds is optically thick in the Lyman continuum, but individual cloudlets are optically thin in the Lyman lines because of the turbulence of the BLR.

\item The reddenings of AGNs implied by observed hydrogen line ratios imply that the typical, nearby, optically-selected AGN is almost an order of magnitude brighter in the far UV.  This resolves the problem of accretion disks being larger than expected for apparent AGN luminosities.

\item Allowance for the greater average extinction in low-luminosity AGNs removes the problem of dust covering factors apparently exceeding 100\%.

\item So-called ``hot-dust-poor" AGNs are simply normal AGNs which happen to have low reddening.

\item The inner most dust attenuating type-1 AGNs originates in a dust driven outflow of the narrow-line region gas from the outer edge of the broad-line region.

\item The extra ionizing radiation from AGNs potentially solves the intergalactic medium ``photon underproduction crisis''.

\item The mass of black holes in the universe is greater than hitherto thought.

\end{enumerate}

\section*{Acknowledgments}

I am most grateful to Ski Antonucci and Bill Mathews for numerous comments and discussions.  I have also benefitted considerably from discussions of hydrogen line ratios with Gary Ferland, and discussions of AGN reddening and dust issues with Victor Oknyansky, Clio Heard, and Jonathan Stern.

%\label{lastpage}

\end{document}